# Machine Learning Prediction of Critical Cooling Rate for Metallic Glasses From Expanded Datasets and Elemental Features

Benjamin T. Afflerbach[a*], Carter Francis[a], Lane E. Schultz[a], Janine Spethson[a], Vanessa Meschke[a], Elliot Strand[a], Logan Ward[b], John H. Perepezko[a], Dan Thoma[a], Paul M. Voyles[a], Izabela Szlufarska[a], Dane Morgan[a*]

[a]University of Wisconsin-Madison, [b]Argonne National Laboratory
[*]Corresponding Authors (bafflerbach@wisc.edu, ddmorgan@wisc.edu)

## Abstract

We use a random forest model to predict the critical cooling rate ($R_C$) for glass formation of various alloys from features of their constituent elements. The random forest model was trained on a database that integrates multiple sources of direct and indirect $R_C$ data for metallic glasses to expand the directly measured $R_C$ database of less than 100 values to a training set of over 2,000 values. The model error on 5-fold cross validation is 0.66 orders of magnitude in K/s. The error on leave out one group cross validation on alloy system groups is 0.59 log units in K/s when the target alloy constituents appear more than 500 times in training data. Using this model, we make predictions for the set of compositions with melt-spun glasses in the database, and for the full set of quaternary alloys that have constituents which appear more than 500 times in training data. These predictions identify a number of potential new bulk metallic glass (BMG) systems for future study, but the model is most useful for identification of alloy systems likely to contain good glass formers, rather than detailed discovery of bulk glass composition regions within known glassy systems.

## Introduction and motivation

Bulk metallic glasses (BMGs) are a class of materials with exceptional properties that support a wide range of application spaces including biomaterials, magnetic devices, and in surface coatings [1,2]. A key challenge in BMG materials discovery is identification of BMG forming compositions in existing glassy alloys and discovery of entirely new BMG alloys. Methods for discovery of BMGs have generally fallen into two broad categories. The first category is qualitative predictions of good glass forming ability (GFA) alloys and regions through identification of various qualitative and semi-quantitative physics-based criteria (e.g., deep eutectics) such as those outlined by Inoue et al [3]. This methodology has had many successes and is responsible for the discovery of many of the BMG alloys known today. The second category is models that quantitatively predict a metric of GFA such as the critical cooling rate ($R_C$) or the critical casting diameter ($D_c$). As our understanding of glassy alloys, and the amount of available data increases, these quantitative models are becoming more appealing as they can potentially reveal much more detailed information about the GFA across alloys.

Quantitative GFA predictions take many forms but can be organized by their choices of features, models, and target predictions. Features typically range from approximately instantly accessible (e.g., elemental properties [4]) to moderately accessible properties needing some calculation (e.g., thermodynamic properties determined from CALPHAD [5], or liquid properties determined by molecular dynamics [6]) to properties requiring extensive synthesis and characterization (e.g., glass transition temperature [7–9] or fragility [10]). Models range from simple linear functions (e.g., the $R_C$ vs. ω correlations [7]) to fully non-linear machine learning models (e.g., $D_c$ vs features fit with boosted trees [9]). Target values range from qualitative categorical predictions (e.g., is a glass under melt spinning [11–13]) to quantitative models of $R_C$ [7] and $D_c$ [9,14–16]. A comprehensive review is not practical here, so we focus on the present status of efforts most similar to ours, where the focus is on instantly accessible elemental property features and quantitative prediction of $R_C$ or $D_c$. We are not presently aware of any study that has successfully built a demonstrably effective predictive model for new BMG systems from simple elemental features. A few notable successes have been the work of Ren et al. and Ward et al., demonstrating a significant ability to predict categorical results of glass forming under melt-spinning, and optimizing GFA of existing known glass formers [17,18]. They fit to over 6,000 melt spinning experiments and achieved a AUC of 0.80 in their ROC curve [17]. Zhang et al. propose a combination of these ideas, using a two-step approach to layer classification predictions with subsequent $D_C$ predictions from a similarly accessible feature set [19]. These works show the power of elemental property features but do not provide an approach to predict new BMG systems. In terms of predicting $R_C$ and $D_C$ there have been striking successes for $R_C$ predictions from characteristic temperatures (liquidus, glass transition, and crystallization temperatures), with Long, et al. reporting an $R^2$ of 0.93 vs. the ω parameter, which is a simple function of characteristic temperatures [7]. $D_C$ has generally been harder to predict quantitatively [7] although Johnson et al. [10] showed an outstanding result $R^2$ value of 0.98 in their predictions for $D_c$ as a linear function of reduced glass transition temperature and fragility. These results suggest $R_C$ is easier to model than $D_C$. These results also suggest that quantitative models of $R_C$ and $D_C$ are possible, although they have only been achieved by using very expensive features that require extensive synthesis and characterization for every new system. However, the above work also suggests that elemental properties can capture physics of GFA, particularly when combined with the ability of modern machine learning methods to model nonlinear relationships and automatically select features. Taken together these observations raise the tantalizing possibility that an accurate model of $R_C$ as a function of elemental features might be achievable.

The absence of a model relating $R_C$ to elemental properties is easily understood as a result of the lack of adequate training data. There are approximately $10^2$ $R_C$ values from direct experimental measurements available. In addition, researcher interest in BMGs and limitations on measuring $R_C$ (typically below $10^4$ K/s) means most data is focused on alloys with known BMGs compositions, and often within composition ranges associated with the BMG formation. A machine learning model that is trained solely on this data will be heavily biased towards predicting that everything is a BMG, limiting the model's utility in identifying new BMG alloys. Limited and biased data are two critical issues holding back machine learning predictions of $R_C$ from simple features like elemental properties. Similar arguments hold for $D_c$, although there are closer to 1,000 data points available [20].

Here we try to develop the first model for $R_C$ as function of elemental features, with a focus on expanding the database of $R_C$ from its directly measured values, as this database is too small to support robust machine learning models. This expansion is accomplished in three steps. First, available $D_c$ data is converted to approximate $R_C$ values using curve fitting to a functional form inspired by simple assumptions about heat transfer during cooling and average thermodynamic properties of metals. Second, available characteristic temperature data is used in combination with previously developed models to estimate $R_C$ for a range of alloys. And third, available melt spinning data is assigned approximate values for $R_C$. The goals of adding these different set of data are to provide more varied compositional space, increase the amount of training data, and expand the range of $R_C$ values available for training. These methods expanded the amount of training data available by over an order of magnitude compared to direct measurements of $R_C$. Using this new dataset, a random forest (RF) model has been trained and evaluated for accuracy in predicting $R_C$ and has also been used to predict the GFA in new BMG systems.

## Database details and Computational methods

### Source Database

The starting $R_C$ database was obtained primarily from Long et al. who gathered 53 experimental measurements of critical cooling rate [7]. One data point (pure nickel) was removed from this database due to being approximated by different methods. 25 more $R_C$ measurements not in Long et al.'s database were found from eight more papers for a total of 77 experimental $R_C$ measurements [21–28]. $R_C$ values are converted to a log scale for easier representation across the wide range of orders of magnitude. Values range from $10^{-2}$ to $10^{7.7}$ K/s with an average of $10^{1.96}$ K/s. We will call this data set 1 (DS1).

### Generated R$_C$ Database

DS1 was expanded three ways. First, we estimated $R_C$ from experimental measurements of critical sizes from casting. We have used measurements of both critical casting diameter $D_C$ and critical casting thickness $Z_C$, and we denote both as $D_C$. Both of these values are converted to $R_C$ values using a generalization of the formalism outlined by Lin and Johnson [29] which suggests the relationship

$$R_C = \frac{A}{D_C{}^B}. \qquad 1$$

Lin and Johnson's proposed equation sets A=10 and B=2 based on assumptions about average thermodynamic properties across all metals and an idealized interface between the alloy and mold. Specifically, they assume a difference between melting temperature and glass transition temperature of 400 K, Thermal conductivity of the melt being 0.1 W/cm s$^{-1}$ K$^{-1}$, and heat capacity per unit volume of 4 J/cm$^3$ K$^{-1}$. B is set to 2 based on an ideal A fit of log($R_C$) vs. log($D_C$) (*Figure 1*) for alloys is DS1 which have both measurements gives A=631 (log(A)=2.81), B=1.8. This fit was then used to approximate $R_C$ values from all $D_C$ and $Z_C$ values without a $R_C$ in DS1. The A and B values shift quite significantly from the values estimated by Lin and Johnson. This difference is likely due to the previous assumptions ignoring surface interactions between the melt and the mold during casting. The application of Eq.1 in this first method added 342 approximate $R_C$ values (which we call Data Set 2 (DS2)) and brings the number of training values up from 77 experimental $R_C$ values to a total of 419 training values.

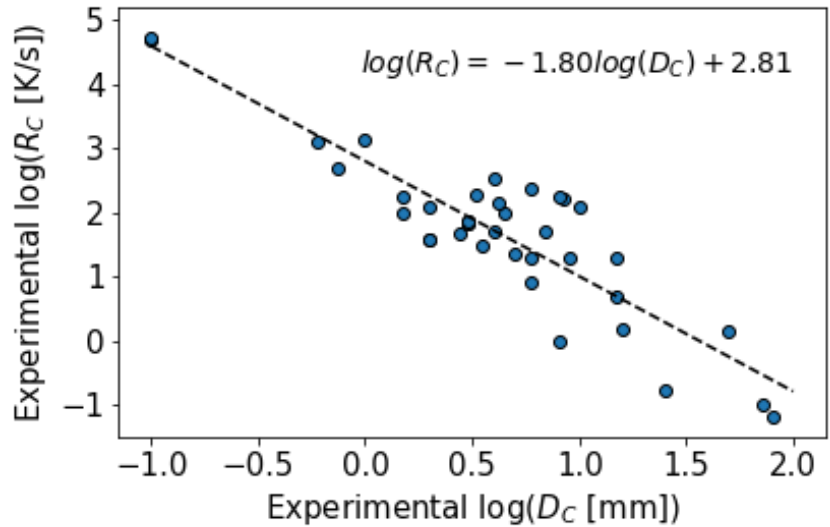

Figure 1. Comparison of a subset of training data with both experimentally measured RC and Dc values. The line of best fit and its equation are shown. The fit has $R^2$ of 0.80, RMSE of 0.55 K/s, and MAE of 0.44 K/s.

Second, we used the ω parameter initially proposed by Long et al to make approximations for $R_C$ for all datapoint for which we have measured $T_g$, $T_x$, and $T_l$ [7]. Specifically, we take all $T_g$, $T_x$, and $T_l$ data we have available, determine ω, and then use the linear relationship between ω and $R_C$ to from Long et al to predict $R_C$. As an additional verification of the ω parameter, for the 25 additional points added to Long et al.'s original data, their ω values were calculated and are shown in the supplementary information as a test set specifically for the ω relationship. All the new values fell within the spread of the previous data, further demonstrating the ability of this parameter to effectively transform characteristic temperatures into estimated critical cooling rate values. Refitting the ω relationship proposed by Long, et al. only resulted in minor changes so to avoid a proliferation of almost identical models we simply used the fitting parameters established by Long et al.. This second method added 141 approximate $R_C$ values (which we call Data Set 3 (DS3)) for compositions that do not overlap with previous datapoints, bringing the total to 560 compositions with approximate $R_C$ values.

Finally, we leveraged melt spinning experiments, which categorize compositions as amorphous, partially amorphous, and crystalline under high-rate cooling. Based on what is known about typical cooling rates during melt spinning, these categories correspond to approximate constraints on $R_C$. Due to the overlap of the expected $R_C$ values for amorphous melt spinning data (such alloys likely have $R_C$ < ~$10^5$ K/s) with significantly higher quality measurements and approximations of $R_C$ from the previous methods, the amorphous category data was excluded from the final dataset. This exclusion is done because introducing such a large amount of very approximate $R_C$ data in the same range where we have access to much higher quality data would likely drown out any signal that would allow the model to differentiate BMGs ($R_C < 10^3$ K/s) from moderate glass formers and non-glass formers. We therefore assigned approximate $R_C$ values only to the partially amorphous and crystalline categories and included them in our fitting. Specifically, we assigned the partially amorphous and crystalline cases $R_C$ values of $10^{5.5}$ and $10^7$ K/s, respectively. When a cooled system comes out partially amorphous it is likely that the actual cooling rate was a little slower than $R_c$, since some of the system had time to crystallize. Furthermore, the cooling rate for melt spinning is known to be in the range $10^4$ and $10^6$ K/s, or based on averaging the logs, about $10^5$ K/s [30]. Therefore, for systems that are partially amorphous it is likely that the true $R_c$ range is somewhat shifted toward higher values than the range $10^4$ and $10^6$ K/s, say $10^{4.5}$ and $10^{6.5}$ K/s. We represent this range by averaging the logs to give $10^{5.5}$ K/s. The value of $10^7$ K/s for the fully crystalline was chosen to be a about one order of magnitude above the fastest cooling rate likely obtained in melt-spinning data to represent the fact none of these alloys formed amorphous structures. The exact $R_c$ value chosen for the crystal forming alloys did not have a significant effect on machine learning performance as we have an extremely small amount of direct experimentally measured $R_C$ values in this range that would be affected by the specific number assigned to this data. Therefore, the main effect of including it and assigning a value is to allow the model to differentiate between the better glass formers found elsewhere in the dataset, and these poor glass formers.

The melt-spinning data is obtained from a review paper which provides over 8,000 melt-spun compositions[31]. From this dataset we used 1248 compositions which formed crystalline metals after melt-spinning, and 720 compositions which were categorized as partially amorphous. Although the $R_C$ values from this data are highly approximate, they are quite distinct from the bulk of the higher-fidelity training data developed above and are therefore expected to constrain the fits without polluting fitting to higher fidelity data. *Figure 2* shows that the crystalline and partially amorphous data do not overlap significantly with the rest of the training data. This process added 1,565 approximate $R_C$ values (which we call Data Set 4 (DS4)) for compositions that do not overlap with previous datapoints, bringing the total to 2,125 compositions with approximate $R_C$ values. This is an increase of almost 30 times greater than the initial set of measured $R_C$ values. We call this final integrated Data Set 5 (DS5).

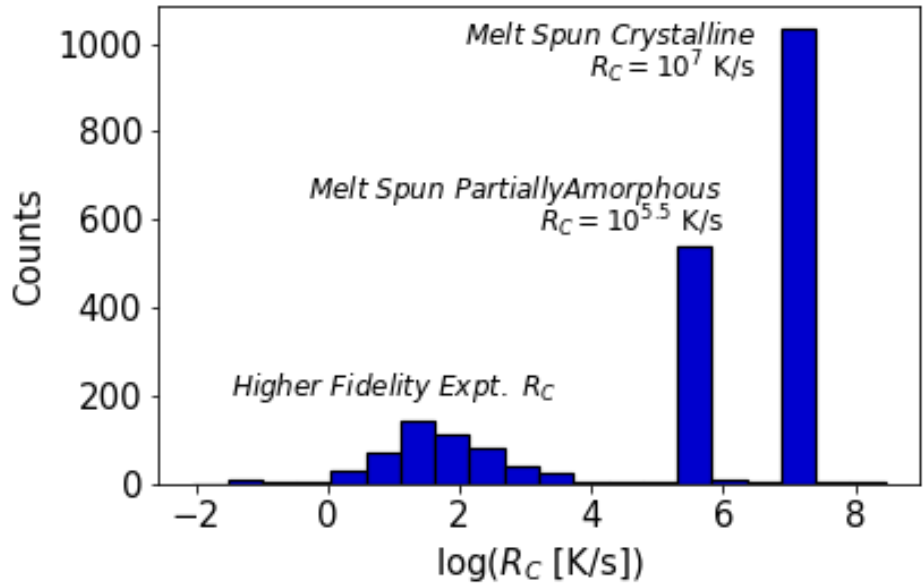

Figure 2. Distribution of $R_C$ values in final training dataset (DS5)

## Machine Learning Models

Using the complete DS5 of $R_C$ data a random forest model was built and trained to predict $R_C$. The random forest model is trained using the MAST-ML machine learning software package which builds machine learning workflows using the underlying scikit-learn python package [32,33]. Inputs to the model are obtained from compositional information and elemental features using the MAGPIE approach proposed by Ward et al. [4,18]. Elemental features for each composition are generated as composition averages, maximum, minimum, and difference. This feature set is chosen to be maximally accessible as all the features can be generated almost instantaneously directly from available elemental databases. Several other model types were also investigated along with the random forest model but showed worse performance. Specifically, gradient boosted trees and Kernel Ridge Regression models showed reduced performance under cross validation testing with a 5-fold cross validated RMSE of 0.732 and 0.803, respectively (compared to 0.36 for random forest, discussed below). We assessed the predictive ability of the model through random and leave-one-group-out cross validation (CV). The random cross validation was done by repeating 5-fold cross validation 10 times (for a total for 50 left folds of data) and the predicted values for each excluded point were averaged.

## Results and Discussion

The average predicted vs. true values are shown in the parity plot in *Figure 3*. We calculated the following statistics from the 5-fold CV test: $R^2$ = 0.97 ± 0.01, root mean squared error (RMSE) = 0.36 ± 0.09, mean absolute error (MAE) = 0.08 ± 0.02, RMSE normalized by the standard deviation of all the log($R_C$) values (RMSE/$\sigma_y$ where $\sigma_y$ = 2.22) = 0.16 ± 0.04 respectively. The error bars represent the standard error in the mean of each statistic when averaged over all 50 CV folds. Although our model uses only elemental features, the errors are comparable to or better than the best previous models for $R_C$ using characteristic temperatures. Specifically, the $\omega$ model for predicting $R_C$ from characteristic temperatures given in Long, et al. showed $R^2$ = 0.90 and RMSE = 0.67 log units [7]. These statistics are influenced by the large amount of melt spinning data which is somewhat unusual due to it being assigned the same $R_C$ value. If the melt spinning data is excluded from the statistics RMSE value increases to 0.70 which is still essentially equivalent to the best previous characteristic temperature based models.

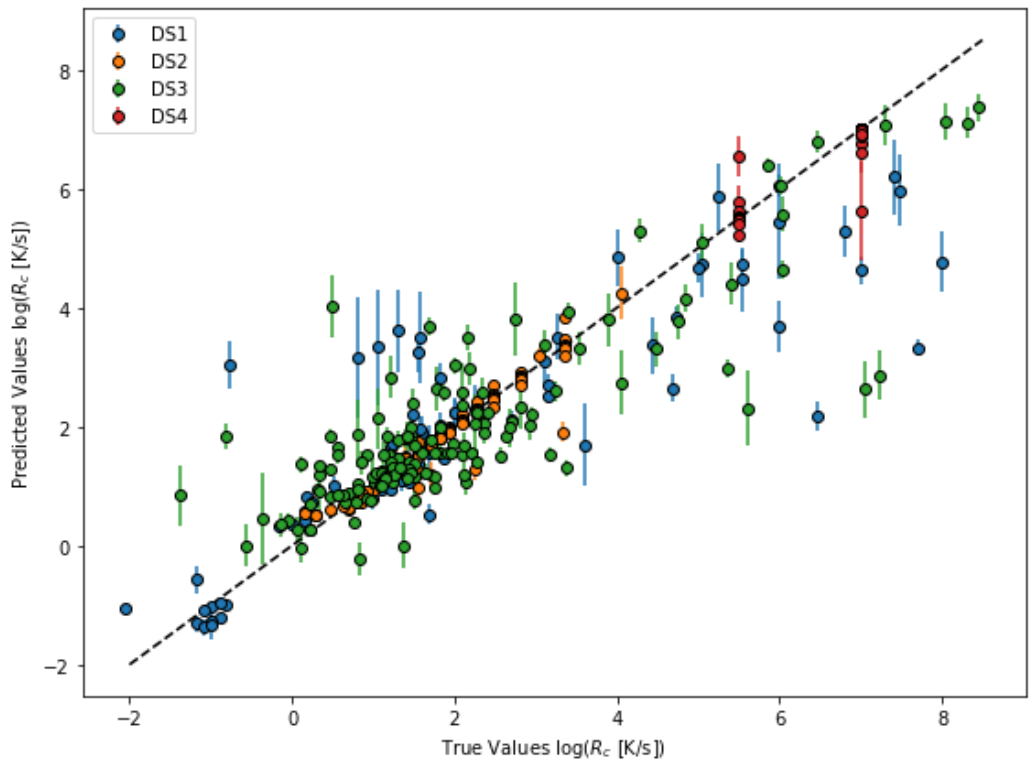

Figure 3. 5-fold cross validation performance of random forest model

While the random CV is a useful standard test of model accuracy it is not a good test of the ability of the model to predict new chemistries [34]. This limitation of the random CV score arises because the data set often has multiple entries on closely related compositions due to the nature of experimental research on GFA, so excluded points in the validation data are likely to have nearby compositions in the training data that contain all the same elements with minor changes to the composition. This allows the model to effectively serve as a "look-up" table for predicting all nearby materials without learning the underlying physics causing good GFA. The random CV score therefore likely overestimates how well the model will perform on new chemistries.

To assess errors on new chemistries we performed a leave-out-one-group CV, where we grouped together similar compositions and left them out one by one, training on the remaining data. Groups were defined by each unique alloy system (i.e., unique combinations of elements). For example, if the dataset only contained three elements the total list of groups would be defined as El1-El2, El1-El3, El2-El3, and El1-El2-El3. As each group was left out the training data the average RMSE were recorded and are summarized in Figure 4. Groups are sorted by the minimum number of overlap instances with the training data between all elements in the group. For example, the elemental overlaps for an excluded Cu-Zr alloy would be the lower value between the number of Cu and Zr containing alloys were left in the training set. The dashed-dotted lines show the average RMSE of all groups within each bin from 0-250, and 250-821 on the x-axis. The dashed line shows the average across all points.

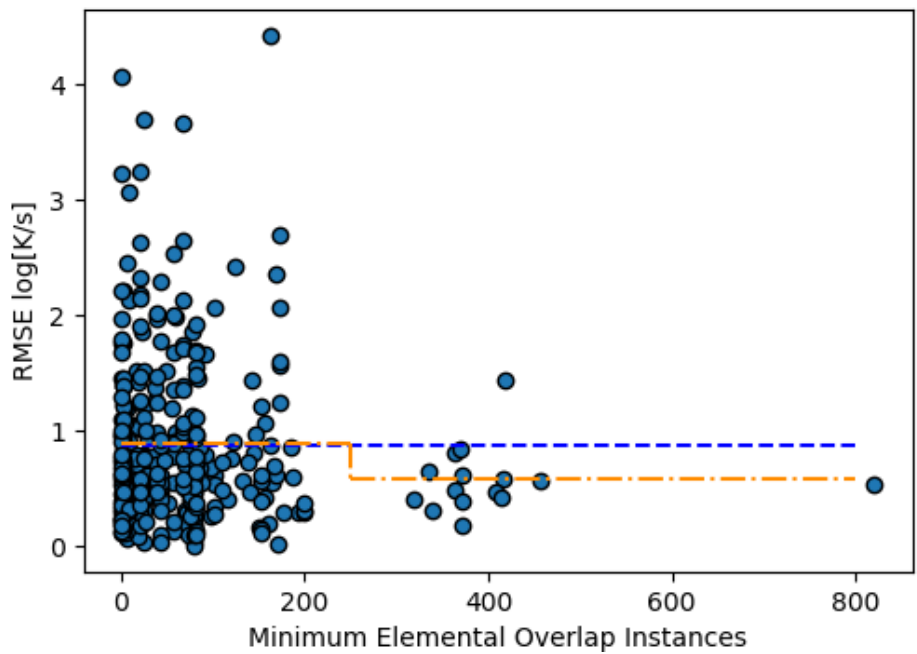

Figure 4. Leave out group cross validation sorted by amount of overlap with training data. The blue dashed line shows the average RMSE of 0.88 log units. The orange dashed-dotted line shows averages for each bin of data from 0-250 (0.89) and 250-821 (0.58).

*Figure 4* shows how the model is performing on average and how it performs when there are many or few representatives of the elements being predicted. The average RMSE (MAE) of 0.88 (0.82) log units is noticeably larger than the random CV RMSE (MAE) of 0.36 (0.08) log units. This increase is due to the larger amount of compositionally similar data being left out when an entire alloy system is removed. Due do the nature of experimental data generation many systems have measurements taken single digit atomic percents away from each other, which may cause the random CV method to overestimate performance. RMSE (MAE) errors still stay below an order of magnitude (one log unit) suggesting that in an average way the model is at least moderately robust to leaving out significant chemical information. One might expect that the model will perform best when there is the most training data. This effect is not particularly apparent from *Figure 4* but the data does seem to cluster into two groups, below about 250 and above, and the RSME goes from 0.89 to 0.58 in going from the low to high group. This result suggests that establishing a cutoff of around 250 elemental overall instances for elements in any predicted alloy systems may help improve the reliability of predictions.

Using the cutoff of 250 instances of overlap we can propose two searches for making predictions with the random forest model to identify new BMG systems. The first search is to use the model to predict likely BMGs from known glass formers from melt-spinning data. As discussed during the database generation section there is a set of melt-spinning data that was left out of model training due to overlap with the higher fidelity experimental data. We will look for BMGs within this group of alloys. We define a BMG as $R_C < 10^3$ K/s. This data has 3,755 compositions that were classified as glass formers under melt-spinning conditions. Of those points there were 63 compositions predicted as $R_C < 10^3$ K/s by our model and therefore predicted to be good BMG candidates. These predictions are shown in *Figure 5.* Predicted critical cooling rates of melt-spun glasses. Points are color coded by interest of the alloy composition. Red points being the least interesting, and yellow points being the more promising as new BMG systems. with more details on the specific alloy systems given in Table 1. The probability of the prediction being a BMG is estimated directly from the random forest confidence interval of each prediction using a one-sided Z-test. An analysis of these estimated confidence intervals is included in the supplementary information in the section Error Bar Analysis.

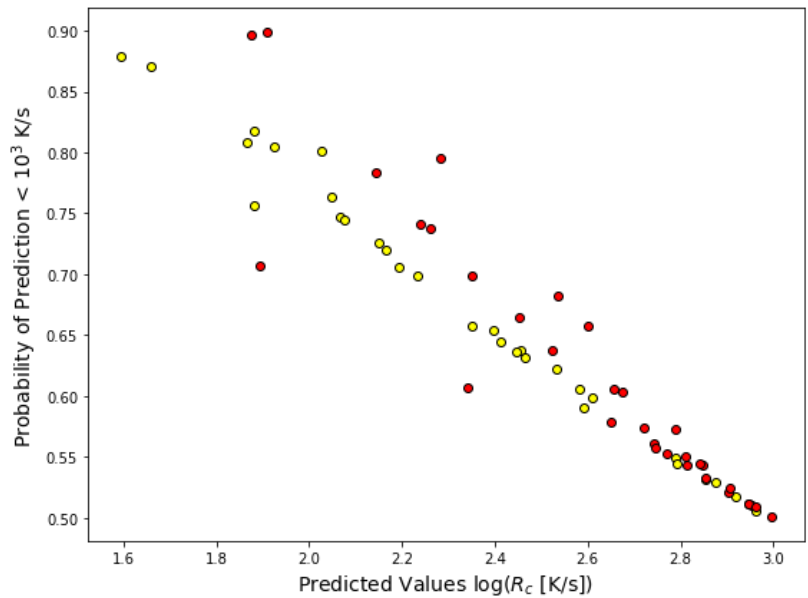

Figure 5. Predicted critical cooling rates of melt-spun glasses. Points are color coded by interest of the alloy composition. Red points being the least interesting, and yellow points being the more promising as new BMG systems.

While the machine learning model can potentially provide helpful guidance in discovering new BMGs, its predictions must be considered by human researchers to assess their value to the community. With this in mind, each prediction in Figure 5 is color coded based on our personal assessment of the novelty of the alloy system for the BMG community. The color scheme is used in Table 1 as well. Red systems are likely the least interesting due to being a known BMG system in our training data. These predictions only demonstrate that the model will predict nearby compositions to training data. Yellow systems are not directly known BMG systems; however they have nearby BMG quaternaries or ternaries with one additional minor alloying element. This limits the novelty of these predictions since we would expect the higher component systems to have better glass forming ability than the predicted lower component alloys. One of these systems, Al-Ca-Ga, is slightly different in that its nearby system is the binary Al-Ca system that is also included in Table 1 which has less components. Of the yellow systems, Al-Ca-Ga is therefore the most potentially interesting as following the same logic this higher number of components in general may increase GFA compared to the known Al-Ca BMG system. Finally, there are several green systems that are potentially the most interesting due to not having any nearby known ternary or quaternary BMG systems. They can all be broadly grouped into the category of Au-B-rare earth. Predictions for these systems fell slightly above the previously established $10^3$ K/s cutoff and are identified with an asterisk. This extension to higher $R_C$ values was considered because previous established estimated errors in predictions still place these systems in the range of being potential glass formers. Based on our literature review this combination of elements appears to be new, with some of the closest systems being the Au-Si-X BMG systems introduced by Schroers et al.[35]. Our predicted alloys essentially replace the Si in the Au-Si-X BMG with another nearby metalloid, B. However, while rare earth elements have been used in BMGs there is not any previous literature combining gold and boron with rare earth elements of which we are aware. Therefore, these types of systems are suspected to be novel and worth additional consideration. As an additional check for potential interest in these systems we consider to what extent they are consistent with previously established empirical rules for finding metallic alloys with high glass forming ability. While these criteria have many forms the following properties of systems proposed by Inoue et al. [36] are generally desirable: (1) multicomponent alloy consisting of more than 3 elements, (2) significantly different size mismatch exceeding 12% among the main 3 constituent elements, (3) negative heats of mixing among their main elements. We also add to this a fourth criteria, which is generally harder to assess without detailed thermodynamics models, which is that the system shows deep eutectics. All alloys are ternaries so do not quite satisfy the first criteria, but we know that BMG ternaries can be formed. All three systems easily satisfy the second criterion due to the large size difference between Boron and the Rare-earth elements. All three systems also satisfy the third criteria. Heats of mixing are calculated for each predicted composition from an extended regular solution model following the methodology and binary interactions from Takeuchi and Inoue [37]. Mismatch percentage along with the estimated heats of mixing, are shown in in Table 1 for the average of predicted compositions in each system. With respect to the fourth criteria, available binary phase diagrams from the ASM Alloy Phase Diagrams Database were analyzed which reveal eutectics in all of the binary subcomponents of the Au-B-X ternary alloys [38]. Specifically, eutectics occur near $Au_{20}B_{80}$, near the edges of the B-X binaries as well as $B_{70}Gd_{30}$, and at many compositions in the Au-X binaries. While we do not have access to full ternary phase diagrams to investigate in more detail, agreement with many previously established criteria for discovery of BMG alloys makes these three systems interesting candidates for further study.

Table 1. List of alloy systems predicted as BMGs. Systems are color coded by potential to be novel BMG (see discussion in text for color coding).

| Alloy System | Known BMG | Size Mismatch | Mixing Enthalpy (Kj/mol) |
|---|---|---|---|
| Cu Hf Nb | No | 22% | -10.1 |
| Cu Nb Zr | No | 23% | -14.3 |
| Cu Ti Zr | Yes | 23% | -16.6 |
| Ni Zr | Yes | 26% | -44.8 |
| Al Ca | Yes | 39% | -18.0 |
| Au B Pr* | No | 131% | -63.2 |
| Cu Hf | No | 22% | -15.9 |
| Al Ca Mg | Yes | 39% | -14.0 |
| Al Co Zr | Yes | 25% | -46.0 |
| Au B Er* | No | 119% | -64.6 |
| Au B Gd* | No | 125% | -64.0 |
| Cu P Zr | No | 32% | -27.4 |
| Al Ca Ga | No | 39% | -22.9 |

To give more insight into the model's predictions, the $R_C$ for systems in Table 1 were predicted over the alloy's full binary and ternary composition ranges and the full predictions for the Au-B-Pr system are shown as an example in Figure 6. We can see from this example that a large portion of the ternary system is predicted near or below the $10^3$ K/s cutoff for predicted bulk formation. This result, combined with the small dynamic range of predictions, with the majority of the ternary predicted within one order of magnitude, suggests we cannot claim to make a prediction of any specific region within the ternary being the most promising. This trend holds across most new predictions, suggesting that for new systems the model predictions can at best identify candidate BMG systems, rather than pinpoint promising BMG regions within systems. This limitation is unfortunate but may be less problematic in the future as the community is developing new combinatorial approaches to experimentally investigate large composition ranges of a system. For example, researchers were recently able to synthesis and characterize $R_C$ over a large region of the Al-Ge-Ni ternary using high-throughput methods [28].

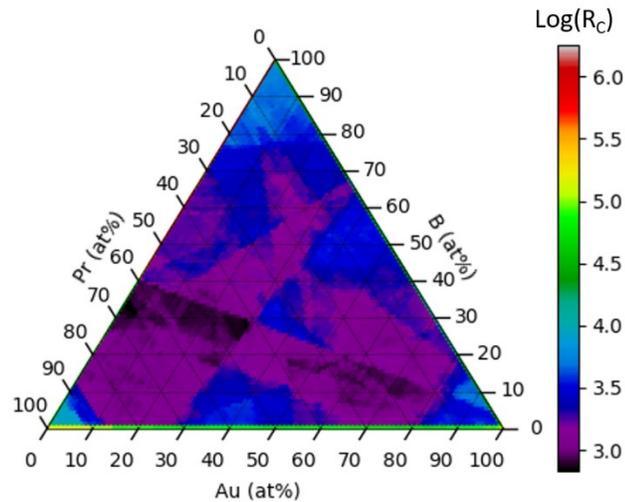

Figure 6. Predictions in 1% composition increments of the Au-B-Pr system.

A second search was performed to explore more widely for potential BMG systems. In the first test above potential systems and compositions were obtained only from known melt-spinning glasses. In this second test we considered every possible quaternary system composed of elements that meet the criterion of more than 250 overlap instances with the training set. There are 10 elements that meet this criterion in the training set: Al, Cu, Ni, Fe, B, Zr, Si, Co, Mg, and Ti. Making every single potential quaternary gave 286 potential quaternary systems. In each system a 10% composition grid was generated for the initial set of predictions. Due to the large number of predictions made multiple steps were taken to filter the predictions to a more manageable number. First all three of the previously discussed criteria proposed by Inoue were filtered against and systems which did not meet the criteria were removed. Notably this means all ternary and binary systems were removed at this stage. Predicted systems were also filtered against known BMG systems in the training data. Predicted compositions were also individually compared to BMG training compositions and removed if they were within 10% elemental composition of any BMG training point. There are 44 systems which meet these criteria, and they are summarized in Table 2.

Table 2. Predictions of GFA for systems constructed from elements with >250 overlap instances in training data. Systems are color coded by potential to be novel BMG (see discussion in text for color coding).

| Idx | Alloy System | Size Mismatch | Minimum Mixing Enthalpy (Kj/mol) | Minimum $R_C$ Prediction log(K/s) |
|---|---|---|---|---|
| 1 | CuFeSiZr | 0.24 | -85.88 | 1.53 |
| 2 | CoCuMgZr | 0.25 | -27.64 | 1.58 |
| 3 | CuMgSiZr | 0.24 | -80.12 | 1.63 |
| 4 | CuFeMgZr | 0.24 | -14.8 | 1.65 |
| 5 | CuNiSiZr | 0.26 | -90.88 | 1.66 |
| 6 | CoCuSiZr | 0.25 | -89.2 | 1.72 |
| 7 | BCuFeZr | 0.81 | -53.88 | 1.86 |
| 8 | BFeMgZr | 0.81 | -49.68 | 1.88 |
| 9 | BCuSiZr | 0.81 | -91.48 | 1.89 |
| 10 | BCuMgZr | 0.81 | -46.04 | 1.89 |
| 11 | BCuTiZr | 0.81 | -60.92 | 1.95 |
| 12 | BCuNiZr | 0.81 | -58.52 | 1.98 |
| 13 | CuMgTiZr | 0.23 | -11.32 | 2.00 |
| 14 | BCoCuZr | 0.81 | -56.4 | 2.04 |
| 15 | CuMgNiZr | 0.26 | -34.2 | 2.13 |
| 16 | BCoFeZr | 0.81 | -61.16 | 2.20 |
| 17 | FeMgSiZr | 0.24 | -82.16 | 2.22 |
| 18 | BFeTiZr | 0.81 | -66.84 | 2.34 |
| 19 | BCuNiTi | 0.65 | -62.24 | 2.36 |
| 20 | CoFeSiZr | 0.25 | -92.36 | 2.40 |
| 21 | CuFeTiZr | 0.24 | -19 | 2.40 |
| 22 | BCuSiTi | 0.65 | -81.2 | 2.43 |
| 23 | BCuMgSi | 0.70 | -47.64 | 2.56 |
| 24 | CoCuTiZr | 0.25 | -28.48 | 2.57 |
| 25 | BCuFeTi | 0.65 | -50.04 | 2.58 |
| 26 | BFeMgSi | 0.70 | -56.28 | 2.60 |
| 27 | BCuMgTi | 0.70 | -51.16 | 2.62 |
| 28 | CoFeMgZr | 0.25 | -28.24 | 2.62 |
| 29 | BFeMgTi | 0.70 | -55.76 | 2.67 |
| 30 | BCoCuTi | 0.65 | -57.24 | 2.67 |
| 31 | CoMgNiZr | 0.26 | -37.48 | 2.68 |
| 32 | BFeSiTi | 0.65 | -80.76 | 2.70 |
| 33 | CoMgSiZr | 0.25 | -85.8 | 2.70 |
| 34 | BCoCuMg | 0.70 | -29.24 | 2.73 |
| 35 | FeMgNiZr | 0.26 | -34.64 | 2.74 |
| 36 | AlCoTiZr | 0.25 | -33.6 | 2.75 |
| 37 | CoCuMgTi | 0.17 | -11.52 | 2.80 |
| 38 | BCuFeMg | 0.70 | -27.8 | 2.81 |
| 39 | FeNiSiZr | 0.26 | -63.04 | 2.82 |
| 40 | BCoFeTi | 0.65 | -62.92 | 2.83 |
| 41 | CoSiTiZr | 0.25 | -94.24 | 2.91 |
| 42 | BCoFeMg | 0.70 | -34.28 | 2.93 |
| 43 | MgNiTiZr | 0.26 | -30.16 | 2.97 |
| 44 | CuFeMgSi | 0.16 | -38.68 | 3.00 |

As with the previous set of predictions we have grouped systems based on relative novelty. We will not discuss all systems in detail but will highlight several trends within the predicted systems as well as commenting on specific systems which may be the most novel. The 16 yellow systems are identified as containing the well-known Cu-Zr binary. Investigating their predictions further showed all predictions had increasing $R_c$ moving away from the Cu-Zr binary suggesting that these alloys are mainly being predicted due to adjacency to the binary. However, they still may be somewhat interesting due to changes in other materials properties while having similar $R_c$. Another trend in predictions is systems that suggest replacements or additions to known ternary or quaternary systems. System 19, B-Cu-Ni-Ti, is somewhat similar to the known Cu-Ni-Ti ternary and Zr-Cu-Ni-Ti quaternary BMGs systems [29,39]. One potential limitation with these systems however is with the combination of B and Ti which, as pointed out by Lin et al. may reduce GFA due to precipitation of very stable borides. They claim reduced GFA in the Zr-Cu-Ni-Ti-B quinary compared to the quaternary without Boron. System 36, Al-Co-Ti-Zr, is somewhat similar to the known Al-Co-Zr system [40], but may provide some different properties. Furthermore, there is a septenary BMG system including all of Al-Co-Ti-Zr elements, further suggesting that these elements may have good glass forming ability [41]. Because Al-Co-Ti-Zr is both a suballoy of a BMG and has suballoys that are BMGs it is a particularly promising system to consider. System 41, Co-Si-Ti-Zr, is one of the more distinct combinations, with no known BMGs in ternaries or quaternaries with simple replacements/additions of single elements. The most similar BMG forming alloy we could identify is reported by Ramasamy et al. in which they replace Nb with Zr In the Fe-Co-B-Si-Nb system to create Fe-Co-B-Si-Zr and report a decrease in GFA due to the replacement [42]. Finally, we identify several systems including Fe-Zr. The Fe-Zr alloy is a well-studied metallic glass though not a BMG [43]. And similar to previous systems there is a known higher component BMG system in the Fe-Co-Ni-Zr-Mo-B system[44]. Systems 19, 36, 41, and the Fe-Zr containing alloys make up all the 7 green systems in Table 2. By predicting across such a wide composition space we identified systems that build off of binary BMGs, proposed substitutions to ternary and quaternary glasses, and also predict entirely knew alloys with no nearby known glass formers. Making predictions with such variety can hopefully inspire new synthesis and discovery of BMG alloys.

With these searches complete we take a step back to analyze in more detail our confidence in the predictions of new BMG compositions. Although our model is formally regression fit to $R_c$, in predicting new BMGs we have effectively used it as a classifier which predicts either BMG or not BMG if the predicted $R_c$ is > $10^3$ K/s or < $10^3$ K/s, respectively. We can therefore ask the classification model question, what is the probability of an alloy actually being a BMG given that the model has predicted it to be a BMG (i.e., what is our precision)? The precision (and recall) can be estimated for our particular data by finding the true positive rate (*TPR*) and false positive rate (*FPR*) from the left-out data in the 5-fold CV tests performed in Figure 3. This yields very encouraging results, with TPR = 0.963 and FPR = 0.013. In addition to this single set of statistics for a single cutoff, we report the full ROC curve in Figure 8 of the supplementary information which has an area under the curve of 0.998. In other words, for alloys left out in a fold, the criteria $R_c$ is < $10^3$ K/s for being a BMG correctly identifies an alloy with a known $R_c$ < $10^3$ K/s 0.963 fraction of the time and finds an alloy with $R_c$ > $10^3$ K/s 0.013 fraction of the time. However, the database used here is quite different from the composition space we explore when we looked at all quaternaries made from 10 elements in the second search above. In particular, the database we are using has far more BMGs than likely in the random search, which changes the probability of correctly identifying a BMG. It is therefore necessary to correct the probability of finding a BMG derived for our database for the fact that BMGs are quite rare in our new search space. We therefore used Bayes Theorem to estimate a more accurate probability of correctly predicting a BMG from the space of relevant systems in our 10 element search. Equation 2 below shows the details of Bayes theorem applied to the present calculation

$$\Pr(BMG|BMG_{pred}) = \frac{TPR*\Pr(BMG)}{TPR*\Pr(BMG)+FPR*(1-\Pr(BMG))}. \qquad 2$$

Here $\Pr(BMG|BMG_{pred})$ is the probability of finding a BMG given that we predict a BMG, which is what we seek, and $\Pr(BMG)$ is the probability of finding a BMG from a random alloy. TPR and FPR are estimated above. $\Pr(BMG)$ is more difficult to obtain so we propose here a few methods. The first 3 methods build from DS5 and count all the BMG datapoints within the dataset. We then identify all elements which compose these BMG alloys, 41 elements total, and define a total composition space of every elemental combination up to quinary alloys in 1% composition increments. Doing this gives 838 BMG alloys in DS5 out of a total compositional space of 3.45×10$^{12}$ potential alloys, for a probability of finding a BMG at random of 2.43×10$^{-10}$. This first method assumes that the 828 BMG alloys in the dataset account for all the actual BMG alloys in this entire compositional space, which is a very pessimistic assumption, and therefore serves as a lower bound on this estimate. Methods 2 and 3 modify this initial estimate as a probability ten times and one-hundred times this to represent possibilities that the current 838 known BMG alloys only comprise 10% or 1% of the actual number due to currently undiscovered alloys which could still be found in a random search. To give an upper bound on this type of analysis we also propose a fourth method taken from an estimation performed by Li et al. in which they performed a theoretical search for bulk glass formers using a number of previously established rules of thumb for identifying BMGs [45]. In their study they estimated about 1% of syntheses of potential glassy alloys results in discovery of a BMG. This 1% estimate therefore represents the probability of randomly discovering a new BMG given that you are an expert researcher using knowledge to pick initially promising materials. Values and results for these four methods are shown in Table 3.

Table 3. Probability estimates and results for a Bayesian analysis of probabilities of finding BMGs.

| Method Number | Probability of Randomly Finding BMG | ML True Positive Rate on Database | ML False Positive Rate on Database | Probability of ML Prediction being BMG |
|---|---|---|---|---|
| 1 | 2.43e-10 | 0.963 | 0.013 | 1.77e-8 |
| 2 | 2.43e-9 | 0.963 | 0.013 | 1.77e-7 |

| 3 | 2.43e-8 | 0.963 | 0.013 | 1.77e-6 |
| 4 | 0.01 | 0.963 | 0.013 | 0.42 |

As noted above, the model has a TPR of 0.963 and a *FPR* of 0.013 on the database we have used for cross validation, which suggests that the trained ML model should be quite good at identifying BMG alloys from data like that used in the cross validation. However, when factoring in the overall very small population of BMG alloys within a likely search space using Eq. (2) above, the probability that any predicted BMG alloy will actually be a BMG when synthesized becomes very low for methods 1, 2, and 3. These probabilities range from an approximate $10^{-8}$ to $10^{-6}$ depending on which the assumption for how many BMGs within the elemental set from DS5 have been found. This result highlights that even with fairly good cross-validated performance statistics, machine learning models are not sufficient for the discovery of new materials if the material is rare in the search space and no human guidance is given. If now we consider method 4, in which we replace our estimate of finding a BMG with that estimated for a search space selected by domain experts, we calculate the probability that our model correctly identifies a new BMG when it predicts one to be 42%. What this result implies is that the machine learning model is likely almost useless for finding BMGs when used on random alloys, but potentially quite useful when used on a set of alloys prescreened by human experts using qualitative rules of thumb. In general, this result suggests that a hybrid approach in which machine learning models are not blindly trusted, but merged with existing domain knowledge and human selection, can massively improve the likelihood of materials discovery.

## Conclusions:

A machine learning model predicting critical cooling rates directly from compositional information was trained and evaluated. The training data for the model was acquired from experiments of varying leveling of fidelity with various approximations being used to combine the data in a single dataset of critical cooling rates. The model shows promising predictive ability in alloys with significant elemental representation in the training data. However, predictive ability where this overlap is low drops off considerably and the likelihood for large errors in predictions increases. Furthermore, predictions of specific composition regions within an alloy system are usually within the uncertainty of predictions which suggests that the model is likely best used for identifying potential BMG systems as opposed to searching within new systems for optimal BMG regions. Viewing the results through the lens of Bayesian statistics demonstrates that although results seem promising the ability for these types of models to reliably predict new BMG models is significantly limited by the overall low likelihood of finding BMGs. Therefore, there is still need for improvements and tight integration with human guidance before machine learning models can be used to rapidly discover new BMGs.

## Data Availability

All datasets and machine learning results which includes data for all figures and tables can be found on FigShare at ([10.6084/m9.figshare.15160197](10.6084/m9.figshare.15160197)). The assembled dataset is also available through the materials data facility (10.18126/nc04-ibut).

## Supporting Information

The Supplementary Information is contained in a single file consisting of five sections:

- Complete Cross Validation Analysis
- Error Bar Analysis
- Comparison of Omega Parameter with Additional Data Points
- Complete ROC Curves for Classification of BMG Predictions
- Analysis of Melt-Spun $R_C$ Value Assignment

## Acknowledgements

The authors gratefully acknowledge support from NSF DMREF award number DMR-1728933.

**For Table of Contents Only**

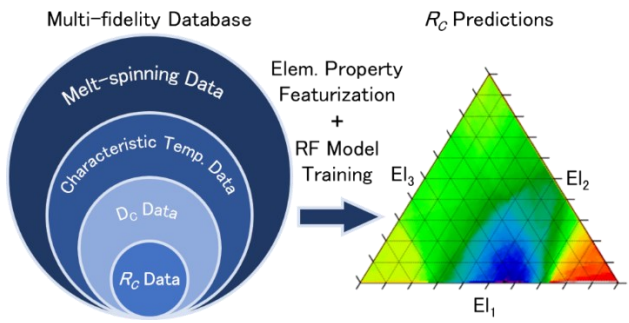